\documentstyle[twocolumn,aps]{revtex}
\begin{document}

\draft

\title{Skyrmions and edge spin excitations in quantum Hall droplets}

\author{J. H. Oaknin$^1$, L. Mart\'{\i}n-Moreno$^2$ and C. Tejedor$^1$}

\address{$^1$Departamento de F\'{\i}sica de la Materia Condensada,
Universidad Aut\'onoma de Madrid, Cantoblanco, 28049, Madrid, Spain.}

\address{$^2$Instituto de Ciencia de Materiales (CSIC),
Universidad Aut\'onoma de Madrid, Cantoblanco, 28049, Madrid, Spain.}

\date{\today}

\maketitle

\begin{abstract}

We present an
analysis of spin-textures  
in Quantum Hall droplets, for filling factors $\nu \simeq 1$. 
Analytical wavefunctions with well defined quantum numbers are given for the 
low-lying states of the system which result to be either bulk skyrmions or 
edge spin excitations. We compute dispersion relations and study how skyrmions 
become ground states of the Quantum Hall droplet at $\nu \gtrsim 1$. 
A Hartree-Fock approximation is recovered and discussed for those spin textures.

\end{abstract}

Spin excitations of a two-dimensional (2D) electron gas in the 
filling factor $\nu =1$ regime are the subject of increasing interest due to 
the experimental observation\cite{barret,schmeller,goldberg} of 
skyrmions.\cite{rezayi,sondhi,fertig,moon,he,brey}
They have been  
analyzed both by a classical nonlinear-$\sigma $ model (CNLSM) 
\cite{sondhi,moon} 
and by a Hartree-Fock (HF) approximation.\cite{fertig,brey} These descriptions  
ignore quantum fluctuations; furthermore  
quantum numbers as modulus ($S^2$) and third component ($S_z$) 
of the spin, and total ($M$) and center-of-mass 
($M_{CM}$) angular momenta are not well described. 
A superposition of mean-field wavefunctions has been proposed\cite{wilczek}
which have well defined $M$ and $S_z$
but still neither $M_{CM}$ nor $S^2$ are represented adequately.  
Current descriptions of spin-textures have  
always been considered in infinite\cite{sondhi,fertig,moon,brey} 
or periodic (sphere)\cite{rezayi,moon,wilczek} systems.  
Therefore, they can not be directly used to analyze 
available experimental information for small systems.\cite{ashoori,klein,haug} 

In this letter we address the two previously mentioned restrictions 
of present theories: 

1) We analyze spin-textures in a system of $N$ electrons moving in 2D,
in the presence of both a high perpendicular magnetic field $B$ 
and a confinement potential (Quantum Hall droplets (QHD)). 

2) We obtain analytical many-body wavefunctions of spin excitations with all 
quantum numbers ($M$, $M_{CM}$, $S^2$ and $S_z$) properly described, 
i.e. we present a microscopic description including quantum fluctuations. 

We also obtain the energies of these excitations, which turn out to
be low-lying excitations at $\nu = 1$, some of them becoming the ground state
(GS) close to $\nu =1$.
 
We consider interacting electrons moving in the $xy$ plane, confined by a 
parabolic potential (characterised by a bare frequency $\omega_0$). 
We assume that 
$B$ is high enough as to project the Hamiltonian H onto 
the lowest Landau level with both spin up and down. In symmetric gauge,   
\begin{eqnarray}
H \! = \! H^{SP} \! + \! \sum_{m_i, \sigma _j} 
\! \frac{V_{m_1m_2m_3m_4}}{2} c^{\dagger}_{m_1,\sigma _1}
\! c^{\dagger}_{m_2,\sigma  _2} \! c_{m_3,\sigma  _2} \! c_{m_4,\sigma _1}
\label{ham}
\end{eqnarray}
where $c^{\dagger}$ and $c$ are the electron creation 
and annihilation operators,
respectively, $m_i$ are single-particle (SP)
angular momenta and $\sigma _j$ label 
spins. We consider matrix elements $ V_{m_1m_2m_3m_4}$ for the Coulomb 
interaction, although we have checked that our conclusions do not depend 
on the details of the electron-electron repulsion.
The single particle part $H^{SP}$ has a spectrum 
\begin{eqnarray}
E^{SP} (M,S_z)\! = \! \left( \frac {N}{2}+\frac{ \Omega -\omega _c} 
{2\Omega }M \right) \hbar \Omega + \! g \mu _B S_z B
\label{hamSP}
\end{eqnarray}
where $\omega _{c}=eB/(m^\ast c)$, $\Omega =[\omega _{c}^{2}+4\omega 
_0^{2}]^{1/2}$, 
$g$ is the Lande factor and $\mu _B$ the Bohr magneton. 
Due to the symmetry of the system
the spectrum separates in subspaces with well defined
${M,M_{CM},S^2,S_z}$. The energy ordering of states within a subspace
does not depend neither on $\omega_0 $ nor on $B$; only the relative
positioning of energies in different subspaces do.

Some of these subspaces have dimension 1, so the state which spans
each one of them is an eigenstate of the many-body Hamiltonian. 
This is the case for single Slater determinants comprised of the $k$ lowest 
$m$ SP states with spin $\downarrow$ and the $N-k$ lowest
$ m$ SP states with spin $\uparrow$, where $k=0,1,...N/2$.
We label these eigenstates as $|C_{N-k}^k\rangle$, and we 
refer to them as "compact" states, due to the the fact that they 
have minimum $M$ compatible with a given $S_z=(N-2k)/2$.
Numerical calculations\cite{palacios} show that compact states are GS
for some ranges of the parameters defining the QHD. For
example, on increasing $B$ the transition from the paramagnetic
$|C_{N/2}^{N/2}\rangle$ ($\nu=2$) to the ferromagnetic  
$|C_{N}^0\rangle$ ($\nu=1$) occurs by flipping spins through all 
$|C_{N-k}^k \rangle$. However, these are not the only GS in the process,
other GS with small $S_z$ appearing in between any spin-flip transition.
\cite{palacios} Small-$S_z$ states appear as well as GS for  
$\nu \lesssim 1$.\cite{eric,kamilla} 

We want to describe these highly correlated low-$S_z$ GS and, more generaly,
low-lying excitations close to $\nu=1$. It has been shown that there is a 
branch of low-lying excitations at $\nu=1$ which resemble magnetoexcitons 
of $|C_{N}^0\rangle$.\cite{oaknin} By analogy, in order to describe 
spin-charge excitations, we try the ansatz of magnon-like excitations:  
${\cal O} ^\dagger _{\pm \Delta
} |C_{N-k}^k \rangle$, with ${\cal O} ^\dagger _{\pm \Delta } =   
\! \sum_{m} o_{m} c_{m\pm \Delta ,\downarrow}^\dagger c_{m,\uparrow}$.
It is known that low-lying excitations have $M_{CM} = 0$.\cite{trugman}
As for charge excitations at $\nu=1$,\cite{oaknin} imposing this condition fixes 
the values of the coefficients $o_m$. We find\cite{else} several possibilities for 
${\cal O} ^\dagger _{\pm \Delta}$ (with $\Delta >0$):

\begin{eqnarray}
\Upsilon ^{\dagger} _{- \Delta } = \! \sum_{m}
\sqrt{\frac{(m+\Delta )!}{m!}}
c_{m - \Delta ,\downarrow}^\dagger c_{m,\uparrow} 
\end{eqnarray} 

\begin{eqnarray}
\Lambda^{\dagger} _{\Delta } \! = 
\! \sum_{m} \sqrt{\frac{m!}{(m+\Delta)!}}
c_{m + \Delta ,\downarrow}^{\dagger} c_{m,\uparrow}
\end{eqnarray}

It can be shown that $({\cal O} ^\dagger _{\pm \Delta})^{n}$, $n$ being
an integer and ${\cal O} ^\dagger _{\pm \Delta}$ any of the previous operators, 
also generate states with well defined $S_z$, $M$, and $M_{CM}$. For that 
purpose, these operators can act on {\it any} state $|C_{N-k}^k \rangle$, except
$(\Lambda^{\dagger} _{\Delta })^n$, which requires $\Delta \leq k$ (note
that therefore none of the $(\Lambda^{\dagger} _{\Delta })^n$ act on the $\nu=1$
state). However, $({\cal O} ^\dagger _{\pm \Delta})^{n}$ 
generate states without a well defined $S^2$. 
Due to the Zeeman term, the lowest energy state in a ($M,S^2$) subspace 
has maximum possible $S_z$. Let us call ${\cal P}$ the operator which 
projects onto the subspace $S^2=S_z(S_z+1)$.\cite{nota0}
So we generate, from compact states, sets of states
$ {\cal P} ({\cal O} ^\dagger _{\pm \Delta})^{n}|C_{N-k}^k \rangle$ 
which have all quantum numbers properly defined. 
In fact, the use of the $ {\cal P}$ operator opens another possibility
for ${\cal O} ^\dagger$-type operators:
\begin{eqnarray}
\Sigma ^\dagger _{1} \! = \! \sum_{m} \sqrt{m+1}
c_{m+1,\downarrow}^\dagger c_{m,\uparrow}.
\end{eqnarray}
acting on any state $|C_{N-k}^k \rangle$ does not have well defined
$M_{CM}$, but ${\cal P} (\Sigma ^\dagger _{1})^n |C_{N-k}^k \rangle$ has
$M_{CM}=0$ for all $n$. 

In order to check whether these ansatz states are eigenstates of
Hamiltonian (\ref{ham}), we have performed numerical 
diagonalisations for $N \leq 9$.
The overlap between 
${\cal P} ({\cal O} ^\dagger _{\pm \Delta})^n |C_{N-k}^k \rangle$
and the exact lowest energy
state in the $(M,S_z)$ subspace they belong to is always higher than $0.99$. 
This is so for ${\cal O}$ being $\Upsilon,\Lambda$, or
$\Sigma $, and for all $\Delta, n,$ and $k$ which satisfy the restrictions
stated above on the applicability to these operators.
There is only one case in which two different states of the form 
${\cal P} ({\cal O} ^\dagger _{\pm \Delta})^n |C_{N-k}^k \rangle$
belong to the same subspace:
${\cal P} (\Lambda ^\dagger _{1})^n |C_{N-k}^k \rangle$
and ${\cal P} (\Sigma ^\dagger _{1})^n |C_{N-k}^k \rangle$
We find that, for $k>0$, the former one has always a high 
overlap with the exact lowest energy state 
of the subspace it belongs to, while
the later one (when orthogonalised to the former one) 
has an overlap again larger than $0.99$ with
one of the low-lying excited states in that subspace. 
For $k=0$ (i.e. $\nu=1$), 
there are not states generated by $\Lambda^{\dagger} _{1}$, and 
$ {\cal P} (\Sigma ^\dagger _{1})^n |C_{N-k}^k \rangle$ becomes 
the lowest energy state of the subspace with $M=N(N-1)/2+n$ and $S_z=N/2-n$.

Now we investigate the physical nature of the different kinds of states. 
Let us focus first on states of the form ${\cal P}(\Lambda ^\dagger _{1})^n
|C_{N-1}^1 \rangle $. Simpler wavefunctions 
can be obtained by the inverse transformation to the one
which projects a skyrmionic BCS-like single Slater 
determinant into a state with well defined $M$.
\cite{moon,wilczek} We obtain 
\begin{eqnarray}
| \Psi(\varphi, \xi)  \rangle & = &
\sum_{n=0}^{N-2} \frac{e^{in\varphi }}{n!}
\xi ^n (\Lambda _1^\dagger )^n |C_{N-1}^1 \rangle   
\label{BCS} \\ 
& = & -c_{0 \downarrow }^\dagger \prod _{j=0}^{N-2}
(c_{j \uparrow }^\dagger + \frac{\xi }{\sqrt {j+1}}
e^{i\varphi } c_{j+1 \downarrow }^\dagger ) |\phi \rangle 
\label{HF}
\end{eqnarray}
$|\phi \rangle $ being the vacuum state. From this form, $| \Psi(\varphi, \xi)  
\rangle$ is readily recognised as a HF wavefunction describing skyrmions 
with topological charge 1;\cite{fertig,moon,brey} it depends on two parameters: 
$\varphi$ is a variable fixing a broken-symmetry 
direction of the spin in the $xy$ plane, 
while parameter $\xi$ controls the skyrmion size (i.e. its
$M$ and $S_z$). From eq. (\ref{HF}), $S_z^{SK}(r)
\sim (r^2-2\xi ^2)/(r^2+2\xi ^2)$ at large distances, as for 
skyrmions in the CNLSM, which implies a 
size\cite{fertig} $\lambda=\xi /\sqrt {2}$. Remarkably, such dependence of
$S_z^{SK}(r)$ holds even in the case of skyrmion sizes comparable with
the magnetic length $l_B=\sqrt {\hbar /m^\ast \Omega }$, regime well 
beyond the range of applicability of the classical model. 
A technical point, of relevance in the interpretation of these states,
is that in eq. (\ref{BCS}) the projector ${\cal P}$ is not
included. Otherwise $| \Psi(\varphi, \xi)  \rangle $ would not be
a BCS-like single Slater determinant. 
Corrections introduced by ${\cal P}$ are
proportional to $S_{z}^{-1}$, so for $n<<N$ and $N>>1$ they are negligible.
In the small $N$ limit, or for $n \sim N/2$, the skyrmion is "deformed" and
its identification as such becomes fuzzier. For instance, we find (for $N$ even) 
${\cal P} (\Lambda ^\dagger _{1})^{(N/2)-1}|C_{N-1}^1 \rangle =
{\cal P} (\Upsilon ^\dagger _{-1})^{N/2}|C_{N}^0 \rangle $.
 
By a similar argument, we identify excitations created by $\Lambda ^\dagger 
_{\Delta}$ as skyrmions with topological charge $\Delta$. 
Recall that, although skyrmions with $\Delta>1$  are never 
the lowest energy excitation branch in this model, they are
eigenstates  appearing from partially-ferromagnetic compact systems with, 
at least, $\Delta$ electrons with spin $\downarrow$. 

A physical identification of states generated by $\Sigma ^\dagger _{1}$ or 
$\Upsilon ^{\dagger} _{- \Delta } $ is not so direct. 
From the form of the operators both of them are {\it edge} spin-textures. 
Unlike what occurs for skyrmions where $S_z$ changes sign as a function of $r$, 
their $S_z(r)$ is always positive.
They resemble spin-waves in that they appear 
as excitations of any compact (partially-ferromagnetic) system.  

In an infinite system, electron-hole symmetry can be 
invoked to obtain other spin-textures.
In particular antiskyrmions 
arise as $(\Lambda ^\dagger _\Delta )$-type 
excitations of states $|C_k^{N \rightarrow \infty } \rangle $ of holes. 
For the antiskyrmion with topological charge $-1$ and 
$S_z=n$, we obtain the wavefunction
$  (\sum _{m=1}^\infty m^{-1/2} c^\dagger _{m-1,\downarrow }
c_{m,\uparrow } )^n c_{0,\uparrow }|C_{N \rightarrow \infty }^0 \rangle$.  
For a finite system, the hole state complementary to
the electron state $c_{0,\uparrow }|C_{N+1}^0\rangle $ 
is not a compact state and it has 
$M_{CM}\neq 0$. In this case, antiskyrmions cannot be built up by 
operators of the form ${\cal O} ^\dagger _{\pm \Delta}$.

Numerical diagonalisations for small $N$ show that,
on increasing $B$, the spin-flip transition 
from $|C_{N-(k+1)}^{k+1}\rangle $
to  $|C_{N-k}^k\rangle $ occurs through a ladder of states of the form
${\cal P}(\Lambda_1^\dagger )^n |C_{N-(k+1)}^{k+1} \rangle $ and
${\cal P}(\Upsilon_{-1}^\dagger )^n |C_{N-k}^k \rangle $. 
In this letter we concentrate on the transition from $|C_{N-1}^1\rangle $ to
$|C_{N}^0\rangle $, which is sketched in Fig. 1.
Energies of skyrmion-like states $|\Psi^{SK}_n \rangle = 
{\cal P}(\Lambda_1^\dagger )^n |C_{N-1}^1 \rangle $ take the form\cite{else}  
\begin{eqnarray}
\frac{\langle \Psi^{SK}_n|H| \Psi^{SK}_n\rangle}
{\langle \Psi^{SK}_n|\Psi^{SK}_n\rangle} \! = \!  
E^{SP}(M,S_z) \! + \! E_0 \!+ \! \alpha _\Lambda n\! +\! \beta _\Lambda \!  
\left( \begin{array}{c} n \\ 2 \end{array} \right)
\label{energ}
\end{eqnarray}
where $E_0$ is the interaction energy of the initial compact state, 
$|C_{N-1}^1 \rangle $ in this case. 
Both $\alpha _\Lambda $ and $\beta _\Lambda $ have analytical
expressions in terms of $V_{m_1m_2m_3m_4}$ which can be calculated, 
by using Wick's theorem.
We find $\alpha _\Lambda <0$ and $\beta _\Lambda >0$, so
skyrmions in a QHD behave as confined bosons 
interacting via a two-body repulsive
interaction. Figure 2 shows the dispersion relation of the skyrmion branch for
$N=30$. Energies for ${\cal P}(\Upsilon_{-1}^\dagger )^n 
|C_{N}^0 \rangle $ have the form of eq.(\ref{energ}), with different 
expressions for $E_0$, $\alpha _\Upsilon $ 
and $\beta _\Upsilon $. We
find that $\beta _\Upsilon $ is negligible for all $N$.
The transition from $|C_{N-1}^1 \rangle $ to $|C_{N}^0 \rangle $ 
presents two cases: 

i) for $N<N_0$, where $N_0$ depends on the Zeeman
energy but is $N_0 \lesssim 10$, 
the transition is through a ladder 
of states with decreasing $S_z$ first,
until it reaches its minimun possible value 
and increasing $S_z$ afterwards, as depicted in Fig. 1. 

ii) for $N>N_0$ the minimum of the skyrmion branch is not at the end of
the branch ($n=N/2 -1$), as shown in Fig. 2 for $N=30$.
The transition is through all the skyrmion states with $n$ less than a 
certain $n^{SK}$, which depends on $g$, 
going directly to $|C_{N}^0 \rangle $ afterwards without 
passing through any of the ${\cal P}(\Upsilon_{-1}^\dagger )^n |C_{N}^0 
\rangle $ states. $n^{SK}$ labels the spin of the largest skyrmion 
which becomes GS when varying $B$. 

The spin of the GS is a measurable quantity which may be controlled 
varying the Zeeman energy by means of a tilted $B$. Figure 3 shows $n^{SK}$ 
as a function of Zeeman energy, calculated from
$\Psi_n^{SK} \rangle$
and the
HF approximation (HF energies can be obtained using eq. (\ref{BCS})). 
The HF approach gives slightly smaller values for 
$n^{SK}$. This is so because, although the
HF dispersion relation is very accurate (see inset of Fig. 3(a)
and note the small energy scale) 
its minimum is shifted to smaller $n$. 

Experimental information on the transition from $|C_{N-1}^1 \rangle $ to
$|C_{N}^0\rangle $ is already available:
single electron capacitance\cite{ashoori} and transport\cite{haug} experiments
in quantum dots have been interpreted\cite{haug,palacios} in terms of highly 
correlated GS's appearing in numerical diagonalizations performed for 
$N \leq 5$. 
Here, we have identified those states as skyrmions and edge-spin
excitations and shown that skyrmions appear  
for all $N$. 

To complete our discussion on skyrmions we analyse the thermodinamic
limit. In the $g=0$ limit, 
CNLSM predicts that skyrmions in an infinite system have smaller
interaction energy than $|C_{N \rightarrow \infty}^1\rangle $ by an amount
$ \Delta E_{int} = - \sqrt{\pi / 32} (e^2/4 \pi \epsilon l_B)$.\cite{sondhi} 
Figure 4 shows $n^{SK}$ at $g=0$ and the interaction energy of the
skyrmion of size $n^{SK}$ as a function of $N$. $n^{SK}$ grows slowly with
$N$, while $ \Delta E_{int}$ tends to the classical value.

From numerical diagonalisations we know that, for small $N$,
\cite{eric,else} only two kind of excitations become the new GS of the
system when increasing $B$: charge magnetoexcitons $J^\dagger
_{\Delta }|C_{N}^0 \rangle $\cite{oaknin} or edge spin excitations ${\cal P}
(\Sigma ^\dagger _{1})^n|C_{N}^0 \rangle $. However, for an infinite 2D system, 
antiskyrmions must be the new GS of the system for $\nu \lesssim 1$. 
Since we do not have analytical expresions for wavefunctions and energies of 
antiskyrmions in the intermediate $N$ regime, we cannot conclude whether
they are  
the new GS in this case. With this caveat in mind, we concentrate on 
charge and edge-spin excitations at $\nu =1$.   
For arbitrary $N$, expressions for magnetoexciton 
energies have been reported.\cite{oaknin,MSS7} We find that edge spin
excitation energies are of the form (\ref{energ}) with negligible 
$\beta _\Sigma $. Therefore, their energies are linear   
in $n$, i.e. in $M$, see Fig. 2 (notice that this 
is equivalent to a dispersion relation which is  
quadratic in linear momentum $k$, 
as occurs for excitations produced by a broken continuous symmetry). 
Which type of excitations has lower energy depends on
both $N$ and Zeeman energy. 
In the $g=0$ limit, we find that the edge spin branch  
has lower energy than the charge excitation branch for $N<N_{min}$ and 
vice versa for $N>N_{min}$ where  
$N_{min} \simeq 100$ for the Coulomb repulsion. On increasing Zeeman
energy, the charge branch is favoured.  
For instance, for GaAs ($g=0.44$), $N_{min} \sim 20$ for $\hbar \omega 
_0=2meV$ while for $\hbar \omega _0=5.4meV$ the charge branch has lower
energy than the spin branch for any number of electrons ($N_{min}=0$).  

We thank L. Brey and A. H. MacDonald 
for useful discussions. Work supported by CICYT of 
Spain under contracts No. MAT 94-0982-C02 and 94-0058-C02.

\begin{figure}
\caption{Sketch of $M$ and $S_z$ of the different states which, in some 
conditions, become GS of the QHD around $\nu =1$. Compact $(\bullet )$,
skyrmions $(\circ )$, edge spin excitations $(\times )$, $(\diamond )$ 
and charge excitations $(\triangle )$ are shown.}
\label{fig1}
\end{figure}

\begin{figure}
\caption{Dispersion relations of 
skyrmions $(\Lambda _1)$, edge spin excitations $(\Upsilon _{-1})$, 
$(\Sigma )$ and charge excitations $(J)$ for $N=30$. 
$\Delta M$ is the excess angular momentum with respect $|C_{N-1}^1 
\rangle $. 
Parameters for $E^{SP}(M,S_z)$ are taken such that 
$|C_N^0 \rangle (\ast )$ and $|C_{N-1}^1 \rangle (\circ )$ 
have the same total energy.} 
\label{fig2}
\end{figure}

\begin{figure}
\caption{Exact (continous line) and HF (dotted line) 
$S_z$ for the largest skyrmion which becomes GS when varying $B$ 
(in the text refered to as $n^{SK}$)
as a function of g for  
$N=30$ and (a) $\hbar \omega_0 = 2meV$, (b)$\hbar \omega_0 = 5.4meV$. 
Inset in panel (a) shows the comparison 
between exact 
(dots) and HF (dotted line)
total energy for skyrmions (in units of $e^2/(4\pi \epsilon
l_B)$).  Parameters for $E^{SP}(M,S_z)$ are taken as in figure 2.}
\label{fig3}
\end{figure}

\begin{figure}
\caption{$n^{SK}$ (dotted line) and interaction energy of the skyrmion of
$S_z = n^{SK}$ with respect to that of $|C_{N-1}^1\rangle$
(continous line) as a function of N, in the $g=0$ limit. $\ast$ marks the
interaction energy for skyrmions given by the CNLSM
in an infinite system.  All energies are in units of $e^2/(4\pi \epsilon
l_B)$.}
\label{fig4}
\end{figure}


\begin{references}
\bibitem{else}Details of the calculation will be published elsewhere.
\bibitem{nota0} ${\cal P}$ is generated from spin-ladder operators 
$S^{+}$ and $S^{-}$.
\end{references}
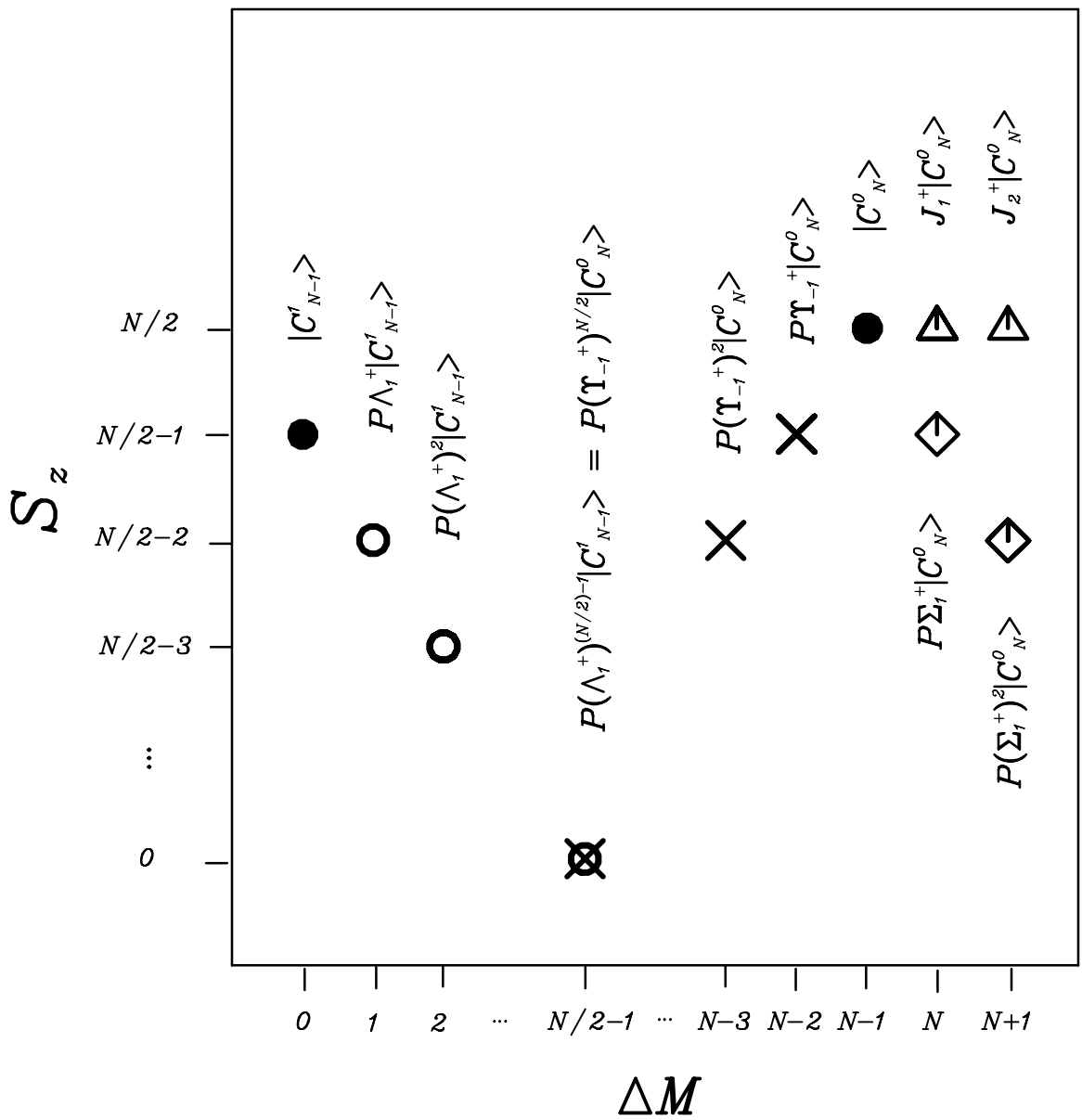
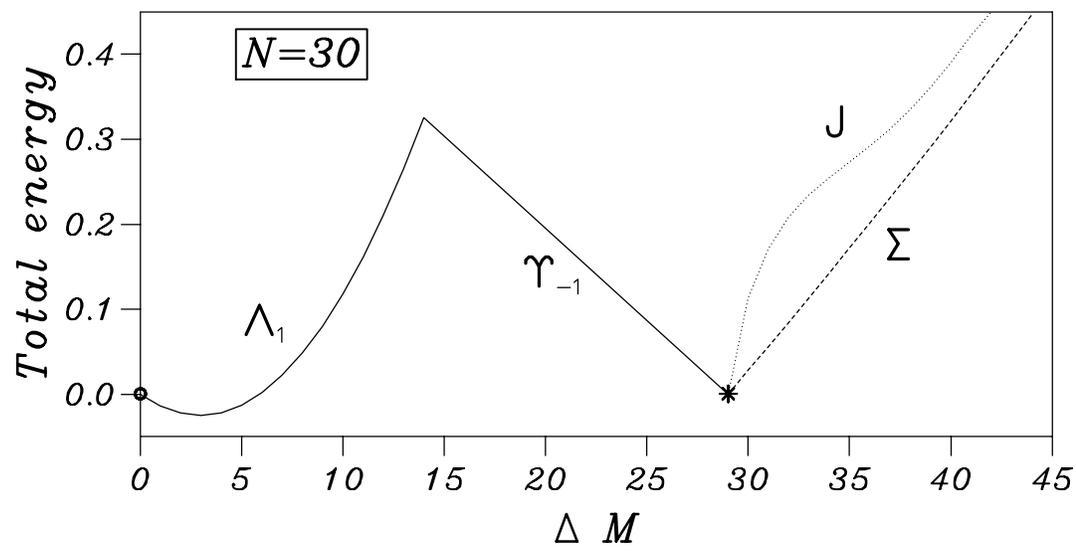
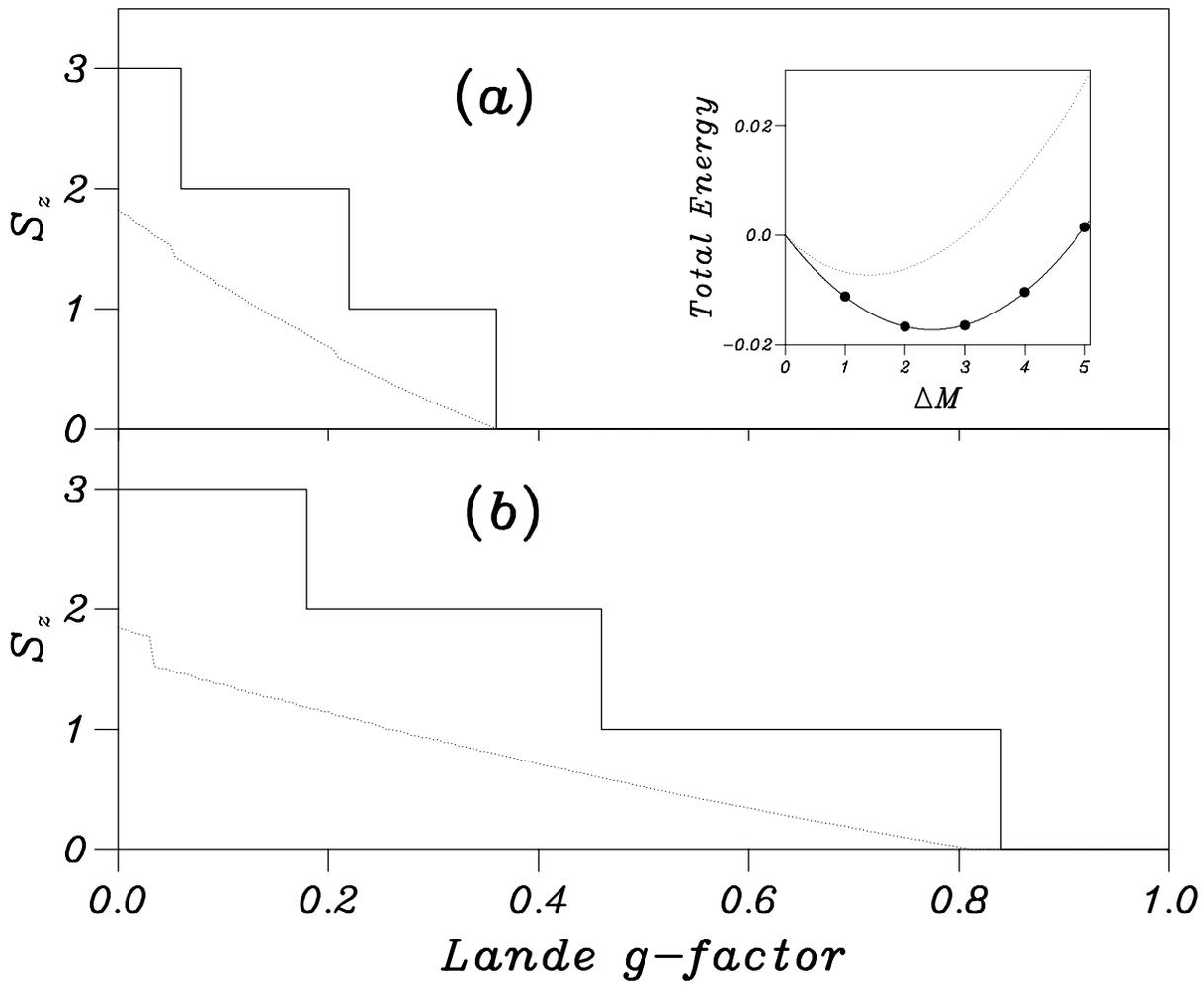
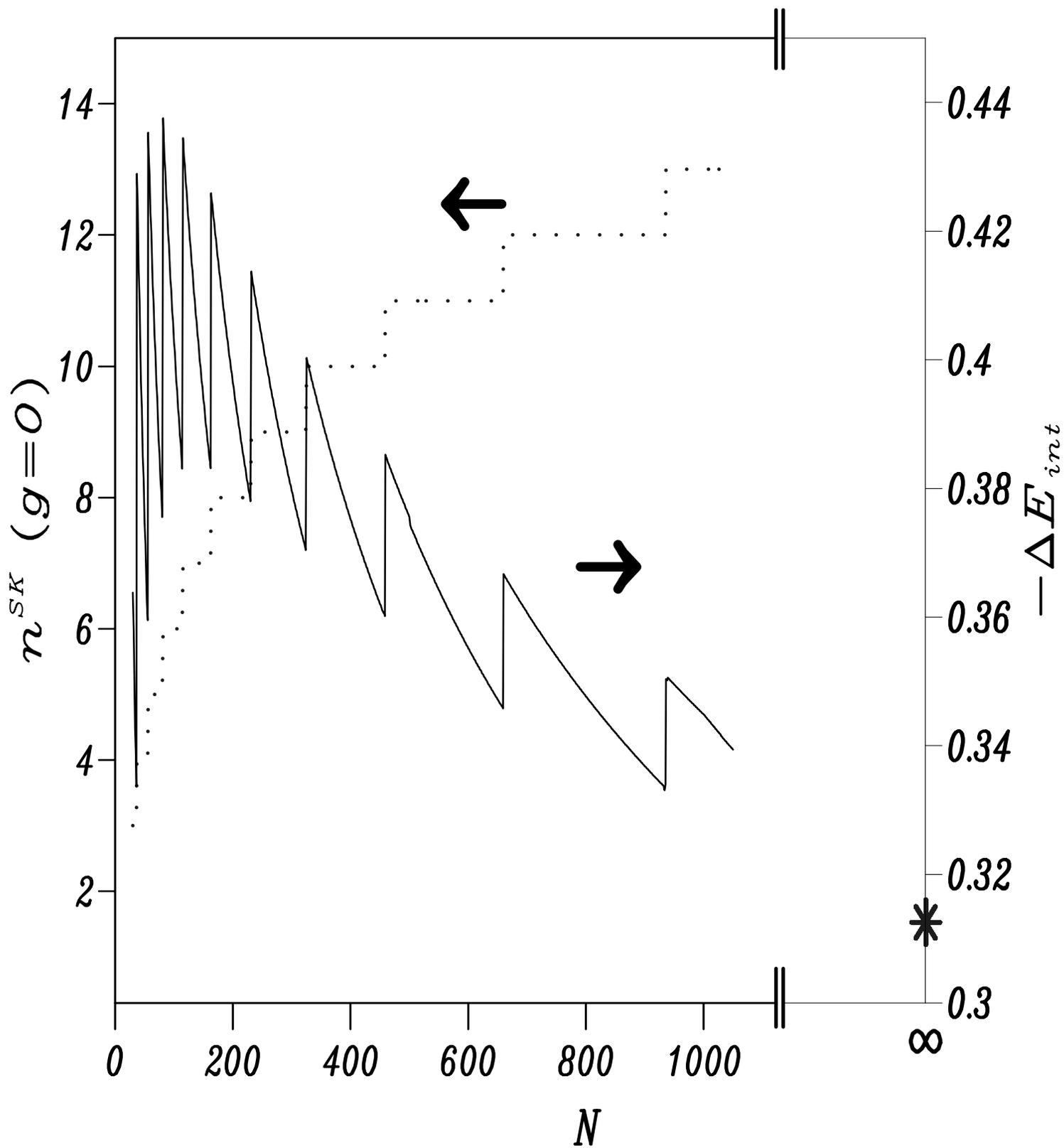
\end{document}